\documentclass[twocolumn, floatfix]{revtex4}
\usepackage{graphicx}
\usepackage{color, url}
\usepackage{amsmath, amsfonts, amssymb, bm}

\begin{document}
\title{Double interatomic Coulombic electron capture induced by 
a single incident electron in two- and three-center atomic systems}
\author{L. M. Ellerbrock}
\author{A. B. Voitkiv}
\author{C. M\"uller}
\affiliation{Institut f\"ur Theoretische Physik I, Heinrich-Heine-Universit\"at D\"usseldorf, Universit\"atsstra{\ss}e 1, 40225 D\"usseldorf, Germany}
\date{\today}
\begin{abstract}
In the process of interatomic Coulombic electron capture, an incident free 
electron is captured at an atomic center $A$ and the transition energy is 
transferred radiationlessly over a rather large distance to a neighboring 
atom $B$ of different species, upon which the latter is ionized. We consider 
two-step cascade processes where the electron emitted from atom $B$ is 
subsequently captured as well, either at center $A$ or yet another atomic 
center $C$, leading to emission of a second electron from one of the centers. 
We derive formulas for the cross section of this double interatomic Coulombic electron 
capture and discuss the relevance of this process, which leads to a substantial rearrangement 
of the electronic configuration, in various two- and three-center atomic systems.
\end{abstract}

\maketitle

\section{Introduction}

Ionization and electron capture processes play an important role 
in various areas of physics, including atomic and molecular physics, 
plasma physics and astro-physics \cite{Review-Recomb2, Review-Recomb3}. 
During ionization, an electron is emitted from a bound state into an unbound 
continuum state, whereas electron capture represents the inverse process.

In some processes, ionization and electron capture occur together.
An example is electron capture from an atomic target by an ionic
projectile in ion-atom collisions: here, an electron transfers from 
an atom to the ion, so that the atom is ionized while the electron is 
simultaneously captured to the ion \cite{coll}. 

Another example, that has been under active scrutiny in recent years,
is interatomic Coulombic electron capture (ICEC). In this process, an 
incident free electron is captured by an atomic center $A$ (which can 
be an atom, ion, quantum dot etc.), transferring the excess energy 
radiationlessly to a neighbouring atomic center $B$, resulting in emission
of an electron from there \cite{ICEC-JPB, ICEC-PRA}. Accordingly, ICEC 
induces effectively a charge exchange between the two atomic centers. 
It requires that the energy set free by the capture transition in 
center $A$ exceeds the ionization potential of center $B$ \cite{2CDR}.
We shall be interested here in ICEC in a system of atomic particles which 
are separated from each other by rather large distances of several angstroms.
Noteworthy, at smaller distances a qualitatively different channel for ICEC,
in which the incident electron just ``triggers'' the process by enabling 
the transfer of an initially bound electron from one atomic particle to another, 
becomes (much) more efficient \cite{ICEC-Sisourat,ICEC-water}.

ICEC has been studied in two-center systems consisting of ions, atoms 
and small molecules \cite{ICEC-JPB, ICEC-PRA}, ions embedded in clusters 
\cite{ICEC-Sisourat}, slow atomic collisions \cite{ICEC-coll}, condensed-matter 
systems of two quantum dots \cite{ICEC-dots1,ICEC-dots2,ICEC-dots3},
and biophysical environments \cite{ICEC-water, ICEC-water2}. While ICEC in 
its original form is a nonresonant process, for certain incident energies 
it can proceed resonantly, when the electron capture at the first center is 
accompanied by the simultaneous excitation of an atomic electron either at 
the same center \cite{res-ICEC1} or the neighbouring center \cite{Remme}.
A review of this research field has recently been given \cite{ICEC-Review}.

Since ICEC is induced by an incident free electron and generates another 
free electron at the end, in principle a series of ICEC events may arise: 
The emitted electron can induce another ICEC in a separate two-center system, 
say $A'$ and $B'$, causing the emission of an electron from $B'$, which can
continue the series. Moreover, it is also conceivable that the original centers 
$A$ and $B$ participate in the second ICEC, which is the subject of the present study.

\begin{figure}[b]
\begin{center}
\includegraphics[width=0.45\textwidth]{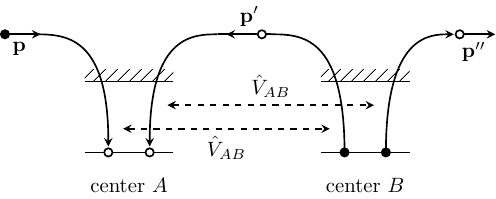}
\caption{Scheme of double ICEC in a two-center atomic system. 
In the first step, an incident electron is captured at center $A$, with the 
excess energy being transferred via interatomic electron-electron correlations 
to atom $B$, which emits an electron. In the second step, the latter electron
is also captured at center $A$ and the transferred excess energy leads to 
emission of a second electron from center $B$.}
\label{fig:scheme2}
\end{center}
\end{figure}

In this paper, we consider processes that consist of two subsequent ICEC 
steps and may accordingly be termed {\it double interatomic Coulombic electron 
capture} (DICEC). In the first step, a free electron is captured at an atomic 
center $A$ and transfers the energy difference to a neighbouring atom $B$ that is 
ionized, ejecting an electron. In the second step, this electron is captured
at center $A$ as well, with the transition energy being transferred again to 
center $B$, from where a second electron is emitted. Thus, in this two-center 
version of double ICEC, two electrons are captured at center $A$, while two 
electrons are emitted from center $B$ (see Fig.~\ref{fig:scheme2}). Symbolically, 
\begin{eqnarray}
e + A^{q_{_A}} + B^{q_{_B}} &\to & A^{q_{_A}\!-1} + B^{q_{_B}\!+1} + e'\nonumber \\
&\to & A^{q_{_A}\!-2} + B^{q_{_B}\!+2} + e''\,,
\label{DICEC_2}
\end{eqnarray}
where $q_{_X}$ indicates the initial charge state of center $X$.

\begin{figure}[t]
\begin{center}
\includegraphics[width=0.47\textwidth]{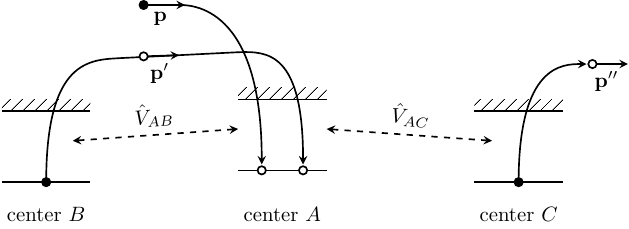}
\caption{Scheme of double ICEC in a three-center atomic system, 
leading to double electron capture at one center. In the first step, as 
in Fig.~\ref{fig:scheme2}, an incident electron is captured at center $A$, 
with the excess energy being transferred to atom $B$, which emits an electron. 
In the second step, the latter electron is captured at center $A$, 
with the transition energy being transferred, however, to another center $C$, 
from where an electron is emitted.}
\label{fig:scheme3b}
\end{center}
\end{figure}

In addition, there is a three-center variant of double ICEC, which also leads 
to double electron capture at center $A$. The transition energy from the 
second capture process is, however, not transfered to center $A$, but rather 
to another atomic center $C$, from where the final electron emission occurs 
(see Fig.~\ref{fig:scheme3b}). That is
\begin{eqnarray}
e + A^{q_{_A}} + B^{q_{_B}} + C^{q_{_C}} \to A^{q_{_A}\!-1} + B^{q_{_B}\!+1} 
+ C^{q_{_C}} + e'\nonumber \\ \to A^{q_{_A}\!-2} + B^{q_{_B}\!+1} + C^{q_{_C}\!+1} + e''.
\label{DICEC_3b}
\end{eqnarray}

\begin{figure}[h]
\begin{center}
\includegraphics[width=0.47\textwidth]{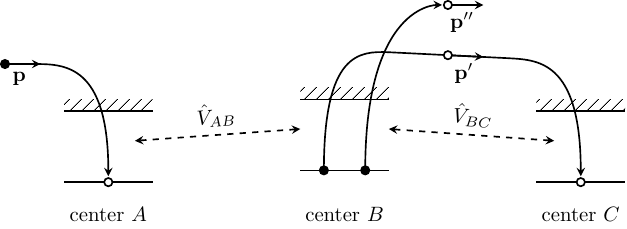}
\caption{Scheme of double ICEC in a three-center atomic system, 
leading to double electron emission from one center. In the first step, as 
in Figs.~\ref{fig:scheme2} and \ref{fig:scheme3b}, an incident electron is captured at center $A$, 
with the excess energy being transferred to atom $B$, which emits an electron. 
In the second step, however, the latter electron is captured at another 
center $C$, and the transition energy is transferred to center $B$, from 
where a second electron is emitted.}
\label{fig:scheme3a}
\end{center}
\end{figure}

There is, moreover, a further three-center variant of double ICEC. Here, 
the electron emitted from center $B$ is not captured at center $A$, but 
instead at a third atomic center $C$, according to (see Fig.~\ref{fig:scheme3a})
\begin{eqnarray}
e + A^{q_{_A}} + B^{q_{_B}} + C^{q_{_C}} \to A^{q_{_A}\!-1} + B^{q_{_B}\!+1} + 
C^{q_{_C}} + e'\nonumber \\ \to A^{q_{_A}\!-1} + B^{q_{_B}\!+2} + C^{q_{_C}\!-1} + e'' .
\label{DICEC_3a}
\end{eqnarray}
The transition energy in the second ICEC step is transferred to the center $B$ 
from where a second electron is emitted. Thus, this version of double ICEC 
leads to double electron emission from center $B$ and involves single electron
capture events at the centers $A$ and $C$.

We note that the DICEC reactions \eqref{DICEC_3b} and \eqref{DICEC_3a} are essentially 
inverse to each other, as becomes apparent by a suitable renaming of the centers.

Our paper is organized as follows. In Sec.~II we present our theoretical 
approach to DICEC in two- and three-center atomic systems. In Sec.~III 
we apply our formalism to various examples and demonstrate the potential 
relevance of DICEC. Our conclusions are given in Sec.~IV. 

Atomic units (a.u.) are used throughout unless explicitly stated otherwise.

\section{Theory of double ICEC}

In this section we present an intuitive approach to double ICEC which 
is based on the known cross section for single ICEC combined with a 
geometrical consideration to account for the second ICEC step. 

\subsection{Cross section for single ICEC}
We start our treatment by introducing the notations and describing
a standard method to calculate the cross section for a single ICEC process. 
To this end, let us consider a system consisting of two atoms $A$ and $B$, 
separated by a sufficiently large distance such that their atomic individuality 
is basically preserved. We note that, while we shall speak throughout of 'atoms'
$A$ and $B$, they could also represent ions, molecules or quantum dots.

Assuming the atoms to be at rest, we take the position of the nucleus of atom $B$  
as the origin and denote the coordinates of the nucleus of atom $A$ by ${\bf R}_{BA}$. 
The coordinates of the electron incident with momentum ${\bf p}$ on atom $A$ 
are ${\bf r}_A = {\bf R}_{BA}
+ \bf{r}$, such that ${\bf r}$ denotes the position vector with respect to the
nucleus of atom $A$. The coordinates of the electron in atom $B$ are called 
$\boldsymbol{\xi}$. The interaction between the electrons reads 
\begin{eqnarray} 
\hat{V}_{AB} = \frac{{\bf r}\cdot \boldsymbol{\xi}}{R_{BA}^3} 
- \frac{ 3 ({\bf r}\cdot{\bf R}_{BA})(\boldsymbol{\xi}\cdot{\bf R}_{BA})}{R_{BA}^5}\ .
\label{V_AB} 
\end{eqnarray}
Since the internuclear distance is assumed to be large, the interaction
$\hat{V}_{AB}$ is small and may be treated perturbatively.
The transition amplitude for the ICEC reaction $e + A + B \to A^- + B^+ + e'$,
where from now on the superscripts indicate the relative change of the charge state,
is accordingly given as 
\begin{eqnarray}
S_{_{\rm ICEC}} &=& 2\pi i \delta(\varepsilon_p + \epsilon_g - \varepsilon_g - \epsilon_{p'}) \nonumber\\
& & \times \langle \varphi_g({\bf r})\chi_{\bf p'}(\boldsymbol{\xi})
|\hat{V}_{AB}|\varphi_{\bf p}({\bf r}) \chi_g({\boldsymbol{\xi}}) \rangle \,.
\label{S}
\end{eqnarray}

Here, the initial state consists of the state $\varphi_{\bf p}$ with energy $\varepsilon_p$
of the free electron incident on atom $A$ and the state $\chi_g$ with energy $\epsilon_g$ of 
the electron bound in atom $B$. In the final state, the initially free electron has been 
captured to the bound state $\varphi_g$ with energy $\varepsilon_g$ in atom $A$, whereas 
from atom $B$ the electron has been ejected into the continuum state $\chi_{\bf p'}$ with 
momentum ${\bf p'}$ and energy $\epsilon_{p'} = \frac{1}{2}{p'}^2$. The latter is determined by 
the energy conservation which is encoded in the $\delta$ function in Eq.~\eqref{S}.

From the transition amplitude one obtains the differential cross section for 
ICEC between $A$ and $B$ as
\begin{eqnarray}
\frac{d\sigma_{_{\rm ICEC}}^{(AB)}}{d\Omega_{\bf p'}} = \frac{1}{4\pi^2}\,\frac{p'}{p}
\,\big|\langle \varphi_g({\bf r})\chi_{\bf p'}(\boldsymbol{\xi})
|\hat{V}_{AB}|\varphi_{\bf p}({\bf r}) \chi_g({\boldsymbol{\xi}}) \rangle\big|^2.
\label{sigma}
\end{eqnarray}
The numerical prefactor arises from the normalization of the continuum states to a unit volume \cite{LL}. The cross section \eqref{sigma} is differential in the solid angle of the electron emitted from atom $B$. This quantity will be needed in our description of double ICEC below.

Before moving on, we note that the total cross section for ICEC can be expressed by 
using the single-center cross sections $\sigma_{_{\rm RR}}^{(A)}$ and 
$\sigma_{_{\rm PI}}^{(B)}$ for radiative recombination of the incident electron 
with center $A$ and for photoionization of center $B$, respectively: 
\begin{eqnarray}
\sigma_{_{\rm ICEC}}^{(AB)} = \frac{\alpha c^4}{\omega_A^4 R_{BA}^6}
\,\sigma_{_{\rm RR}}^{(A)}(\varepsilon_p)\,\sigma_{_{\rm PI}}^{(B)}(\omega_A)\,.
\label{sigma-ICEC}
\end{eqnarray}
This factorization of the ICEC cross section is physically intuitive since the process
involves an electron capture to center $A$ and an electron removal from center $B$.
Here, $\omega_A = \varepsilon_p-\varepsilon_g$ denotes the energy transfer from center 
$A$ to $B$. The numerical prefactor reads $\alpha = \frac{3}{2\pi}$, when the internuclear 
separation vector is aligned with the incident electron momentum, whereas $\alpha = \frac{3}{4\pi}$, 
when an average over the internuclear orientations is taken \cite{ICEC-JPB, ICEC-Review}. 

\subsection{Double ICEC in a two-center system}
In order to theoretically describe double ICEC we shall adopt to the present 
situation a geometrical approach, that was developed in Ref.~\cite{Najjari-2021} 
to treat photo-induced fragmentation of large dimers into singly charged ions. 
In our case, the geometrical approach will be valid when the (reduced) 
de Broglie wavelength $\lambda_{p'}=1/p'$ of the electron, which travels 
between the atomic centers, is substantially smaller than the interatomic 
distance between them. We note that the applicability condition of the geometrical 
approach is perfectly in line with our general assumption that we solely consider 
ICEC when the atoms are separated by large distances.

The first ICEC step between the atoms $A$ and $B$ is characterized by the cross 
section $\frac{d\sigma_{_{\rm ICEC}}}{d\Omega_{\bf p'}}$ from Eq.~\eqref{sigma}
which is differential in the solid angle of the momentum ${\bf p'}$ of the emitted 
electron; the absolute value of this momentum follows from the relation 
$\epsilon_{p'} = \varepsilon_p + \epsilon_g - \varepsilon_g$.
Provided that the energy $\epsilon_{p'}$ of the electron emitted 
from center $B$ is large enough, it can induce a second ICEC between 
$A^-$ and $B^+$. The cross section for the corresponding double ICEC, 
where the emitted electron is captured into $A^-$ while another electron 
is ejected from $B^+$, can be estimated as
\begin{eqnarray}
\sigma_{_{\rm DICEC}}^{(AB)} = \int_{\Delta\Omega_{\bf p'}} 
\frac{d\sigma_{_{\rm ICEC}}^{(AB)}}{d\Omega_{\bf p'}}\,d\Omega_{\bf p'}
\end{eqnarray}
where the integration runs over the element $\Delta\Omega_{\bf p'}$ of the solid
angle of the emitted electron. The latter depends on the internuclear vector 
${\bf R}_{BA}$ and is supposed to have the property that the electron, 
which moves within it, is captured by atom $A^-$ with the probability equal 
to 1, whereas otherwise the capture probability is zero. Since the distance between 
the atoms is very large, $\Delta\Omega_{\bf p'}$ has to be quite small, so that 
it describes geometrically a narrow cone around the internuclear vector 
${\bf R}_{BA}$. Accordingly, one can estimate $\Delta\Omega_{\bf p'}$ as
\begin{eqnarray}
\Delta\Omega_{\bf p'} = \frac{\sigma_{_{\rm ICEC}}^{(A^-B^+)}}{R_{BA}^2}
\end{eqnarray}
where $\sigma_{_{\rm ICEC}}^{(A^-B^+)}$ is the total cross section for the 
ICEC between atoms $A^-$ and $B^+$ (leading to $A^{2-}$ and $B^{2+}$) induced 
by an incident electron of energy $\epsilon_{p'}$. Since the solid angle 
$\Delta\Omega_{\bf p'}$ is very small we can assume that within this angle 
the value of $d\sigma_{_{\rm ICEC}}^{(AB)}/d\Omega_{\bf p'}$ remains 
essentially a constant, that may be taken as 
$\frac{d\sigma_{_{\rm ICEC}}^{(AB)}}{d\Omega_{\bf p'}} = \frac{d\sigma_{_{\rm ICEC}}^{(AB)}}{d\Omega_{{\bf R}_{BA}}}$, i.e. as its value along the direction ${\bf R}_{BA}$.
Then we obtain
\begin{eqnarray}
\sigma_{_{\rm DICEC}}^{(AB)}({\bf R}_{BA}) \approx 
\frac{d\sigma_{_{\rm ICEC}}^{(AB)}}{d\Omega_{{\bf R}_{BA}}}\,
\frac{\sigma_{_{\rm ICEC}}^{(A^-B^+)}}{R_{BA}^2}\ .
\label{sigma-DICEC2}
\end{eqnarray}
The last expression assumes a fixed direction of the internuclear axis 
(with respect to the direction of the incident electron ${\bf p}$). If 
instead the orientation is statistically distributed, we need to take 
an average over the directions, leading to 
\begin{eqnarray}
\sigma_{_{\rm DICEC}}^{(AB)}(R_{BA}) &\approx& \frac{1}{4\pi} \int d\Omega_{{\bf R}_{BA}} 
\frac{d\sigma_{_{\rm ICEC}}^{(AB)}}{d\Omega_{{\bf R}_{BA}}}\,
\frac{\sigma_{_{\rm ICEC}}^{(A^-B^+)}}{R_{BA}^2} \nonumber\\
 &=& \frac{\sigma_{_{\rm ICEC}}^{(AB)}\ \sigma_{_{\rm ICEC}}^{(A^-B^+)}}{4\pi R_{BA}^2}\ .
\label{sigma-av}
\end{eqnarray}

The maximum value in Eq.~\eqref{sigma-DICEC2} is achieved, when the 
internuclear axis lies parallelly to the incident electron momentum
(corresponding to a vanishing polar angle $\vartheta_{{\bf R}}=0$),
because $d\sigma_{_{\rm ICEC}}^{(AB)}/d\Omega_{\bf R}\propto 
1 + 3\cos^2\!\vartheta_{\bf R}$. Accordingly, the maximum value of 
$\sigma_{_{\rm DICEC}}^{(AB)}({\bf R}_{BA})$ is twice as large as 
the averaged cross section in Eq.~\eqref{sigma-av}.

Equation~\eqref{sigma-av} allows for a very intuitive interpretation.
Since the process consistes of two ICEC steps, it is very natural
that the product of the corresponding cross sections appears. The 
factor $\sigma_{_{\rm ICEC}}^{(AB)}$ characterizes the first ICEC step. 
Afterwards, the electron emitted from atom $B$ must travel in the right 
direction in order to hit atom $A$. This is expressed in Eq.~\eqref{sigma-av} 
by the ratio $\sigma_{_{\rm ICEC}}^{(A^-B^+)}/(4\pi R_{BA}^2)$ 
between the cross section of the second ICEC step and the surface of a 
sphere of radius $R_{BA}$ around atom $B$. This ratio measures the probability 
that the emitted electron hits the right spot around atom $A$ of area 
$\sigma_{_{\rm ICEC}}^{(A^-B^+)}$, which represents a very small fraction 
of the full spherical surface around atom $B$. We point out that, as a 
consequence, the cross section of double ICEC scales with the interatomic 
distance as $~R_{BA}^{-14}$, which is steeper than the $R_{BA}^{-12}$ scaling 
stemming solely from the product of the two ICEC cross sections.

\subsection{Double ICEC in a three-center system}
As described in the introduction, there are also three-center variants
of double ICEC, where a third center $C$ is involved [see Eqs.~\eqref{DICEC_3b}
and \eqref{DICEC_3a}]. Let ${\bf R}_{BC}$ be the separation 
vector between the nuclei of atoms $B$ and $C$, and accordingly 
${\bf R}_{AC} = {\bf R}_{BC}-{\bf R}_{BA}$ the separation
vector between the nuclei of atoms $A$ and $C$.

When in the second ICEC step the electron emitted from center $B$ 
is captured at center $A$ with electron emission from center $C$, 
the DICEC cross section is given by
\begin{eqnarray}
\sigma_{_{\rm DICEC}}^{(ABC^+)}({\bf R}_{BA}, {\bf R}_{AC}) \approx 
\frac{d\sigma_{_{\rm ICEC}}^{(AB)}}{d\Omega_{{\bf R}_{BA}}}\,
\frac{\sigma_{_{\rm ICEC}}^{(A^-C)}}{R_{BA}^2}\, .
\label{sigma-DICEC3b}
\end{eqnarray}

When, instead, in the second ICEC step the electron emitted from center $B$ is captured 
at center $C$ with emission of a further electron from center $B$, the DICEC 
cross section reads 
\begin{eqnarray}
\sigma_{_{\rm DICEC}}^{(ABC^-)}({\bf R}_{BA}, {\bf R}_{BC}) \approx 
\frac{d\sigma_{_{\rm ICEC}}^{(AB)}}{d\Omega_{{\bf R}_{BC}}}\,
\frac{\sigma_{_{\rm ICEC}}^{(CB^+)}}{R_{BC}^2}\,.
\label{sigma-DICEC3a}
\end{eqnarray}

Based on Eqs.~\eqref{sigma-DICEC2} - \eqref{sigma-DICEC3a} we can roughly 
estimate the cross section for DICEC. Assuming interatomic distances
of 5\,\AA\ and taking into account that cross sections for single ICEC up to 
$\sigma_{_{\rm ICEC}}\sim 1$\,Mb have been reported in the literature \cite{ICEC-Review},
we obtain $\sigma_{_{\rm DICEC}} \lesssim 1\,$\,kb.

At first sight, the magnitude of the DICEC cross section appears quite small
on the scale which is typical for processes in atomic physics. One has to take
into account, though, that DICEC leads to a very strong rearrangement of the 
electron configuration in the system. Each of the reactions \eqref{DICEC_2} - \eqref{DICEC_3a}
involves three active electrons and four free $\leftrightarrow$ bound transitions,
resulting in changes of the charge state by four units. In light of this, the
magnitude of the DICEC cross section does not appear that small anymore. 
Moreover, DICEC may also include resonant ICEC steps \cite{res-ICEC1, Remme}, 
which can reach cross sections significantly above 1\,Mb at internuclear distances 
of 5\,\AA. In addition, we emphasize that DICEC is extremely sensitive to the 
internuclear distances and grows very quickly when these distances are reduced.

\section{Numerical examples}

In the following, we discuss some examples of atomic systems where DICEC can occur.
Since ICEC benefits from small energy transfers [see Eq.~\eqref{sigma-ICEC}],
our examples involve earth alkaline dications (like Ca$^{2+}$ and Mg$^{2+}$).
They have also been considered in previous studies on ICEC \cite{ICEC-JPB,ICEC-water2}.
Earth alkaline atoms are distinguished by relatively low second
ionization potentials. This can be advantageous because DICEC 
involves capture and/or emission of two electrons from an atomic center.\\

\subsection{Double ICEC in two-center systems}
We start with DICEC in a two-center system, as described by Eq.~\eqref{DICEC_2}.
We consider a combination of two alkaline earth metals, according to 
\begin{eqnarray}
e + {\rm Ca}^{2+} + {\rm Mg} &\to & {\rm Ca}^+ + {\rm Mg}^+ + e' \nonumber\\
&\to & {\rm Ca} + {\rm Mg}^{2+} + e''\,.
\label{DICEC-CaMg}
\end{eqnarray}
The first (second) ionization potential of Ca amounts to 6.11 eV (11.87 eV),
whereas for Mg the first (second) ionization potential is 7.65 eV (14.74 eV)
\cite{NIST}. Accordingly, the incident electron energy must be 
$\varepsilon_p\ge 4.41$ eV, corresponding to a momentum $p\ge 0.57$\,a.u. 

To estimate the DICEC cross section in this system, we first determine the 
cross section for the first ICEC between {\rm Ca}$^{2+}$ and Mg by virtue of
Eq.~\eqref{sigma-ICEC}. The cross section for radiative recombination with 
Ca$^{2+}$ of an electron with incident energy $\varepsilon_p$ can be obtained
from the cross section for photoionization of Ca$^{+}$ (which is available 
in the literature) via the principle of
detailed balance: $g_{_{\rm RR}}p^2\sigma_{_{\rm RR}}(\varepsilon_p) = 
g_{_{\rm PI}}k^2\sigma_{_{\rm PI}}(\omega)$, where $g_{_{\rm RR}}$
($g_{_{\rm PI}}$) is the statistical weight of the quantum states involved 
in the radiative recombination (photoionization) process and 
$\omega = \varepsilon_p + |\varepsilon_g|$ is the photon energy with associated 
photon momentum $k=\frac{\omega}{c}$. Assuming an incident electron energy 
close to the above mentioned threshold and using 
the photoionization cross section $\sigma_{_{\rm PI}}\approx 450$\,kb
of Ca$^+$ from \cite{PI-CaII}, we obtain $\sigma_{_{\rm RR}}\approx 0.05$\,kb
(which is quite small because $k\approx 4.4\times 10^{-3}$\,a.u.).
The photoionization cross section of Mg at about 16.28 eV is 
$\sigma_{_{\rm PI}}\approx 100$\,kb \cite{PI-MgI}.
Accordingly, the cross section for the first ICEC step amounts to
$\sigma_{_{\rm ICEC}}^{({\rm Ca}^{2+}{\rm Mg})} \approx 9/R_{BA}^6$ (in a.u.).
The electron is emitted from Mg with energy $\varepsilon_{p'}\approx   
8.63$\,eV, corresponding to a de Broglie wavelength of
$\lambda_{p'}\approx 1.3$\,a.u.

In the second ICEC step, the electron is captured by Ca$^+$, leading
to emission of another electron from Mg$^+$. The cross section for 
this process amounts to $\sigma_{_{\rm ICEC}}^{({\rm Ca}^+{\rm Mg}^+)}
\approx 3/R_{BA}^6$ (in a.u.). It is obtained from the cross section for 
radiative recombination with Ca$^+$ of about $\sigma_{_{\rm RR}}\approx 
0.004$\,kb (which follows from the corresponding photoionization cross 
section $\sigma_{_{\rm PI}}\approx 350$\,kb of Ca \cite{PI-CaI-1}
and which, again, is small because $k'/p'\approx 5\times 10^{-3}$)
and the photoionization cross section of Mg$^+$ at threshold of
$\sigma_{_{\rm PI}}\approx 270$\,kb \cite{PI-MgII-2}.

The cross section for the double ICEC process in Eq.~\eqref{DICEC-CaMg}
thus becomes 
$\sigma_{_{\rm DICEC}}^{({\rm CaMg})}\approx 9\cdot 3/(2\pi R_{BA}^{14})$,
which which lies below 1\,mb at interatomic distances of $R_{BA}\gtrsim 7$\,a.u.

The reason for the small DICEC cross section in this two-center example 
lies mainly in the smallness of the involved cross sections for radiative 
recombination. While single-center photoionization cross sections are typically 
of the order 0.1--10\,Mb, cross sections for radiative recombination are often 
much smaller, which tends to suppress the magnitude of ICEC [see Eq.~\eqref{sigma-ICEC}]. 
In light of the relationship for detailed balance, this can be understood
by noting that the momentum of a photon with energy, say, $\omega\sim 10$\,eV
is much smaller than the momentum of a few-eV electron. The situation can 
change for photoionization very close to threshold,
where the energy of the ejected electron \cite{remark} is much smaller than the 
ionizing photon energy, so that $p$ is not necessarily much larger than $k$.
This situation, however, cannot be achieved simultaneously for both ICEC steps 
in a DICEC process like in Eq.~\eqref{DICEC-CaMg}, because the first step 
involves the second ionization potential of the atomic species at center $A$ 
and the -- typically much smaller -- first ionization potential of the 
atomic species at center $B$. In the second ICEC step it is vice versa, 
involving the first ionization potential at $A$ and the second ionization 
potential at $B$. Therefore, the various single-center capture and ionization
processes will occur several eV above the respective thresholds. 
As we shall see in Sec.~III.B below, the situation for 
DICEC in three-center systems can be more beneficial.

\medskip
We note that DICEC in a Ca-Mg system could also proceed in inverse order
according to $e + {\rm Mg}^{2+} + {\rm Ca} \to {\rm Mg}^+ + {\rm Ca}^+ + e' 
\to {\rm Mg} + {\rm Ca}^{2+} + e''$. The incident electron energy can be 
arbitrarily small in this case. Exactly at the threshold ($\varepsilon_p=0$)
the cross section of the first ICEC step and, consequently, the cross 
section for DICEC in this system, will diverge. This is because the 
single-center cross section for radiative recombination with the Mg$^{2+}$ 
as a positively charged ion tends to infinity for $p\to 0$. In such a
situation, it would be meaningful to consider as finite quantity the ratio 
$\sigma_{_{\rm ICEC}}^{(AB)}/\sigma_{_{\rm RR}}^{(A)}$ which quantifies 
the relative importance of ICEC compared with radiative recombination \cite{ICEC-Review}.

\subsection{Double ICEC in three-center systems}
We now turn to DICEC in three-center systems. Our first example
considers an earth alkaline dication in the neighbourhood of 
carbon atoms. In such a system, DICEC could process in the 
following way [see Eq.~\eqref{DICEC_3b}]:
\begin{eqnarray}
e + {\rm Mg}^{2+} + {\rm C} + {\rm C} 
&\to & {\rm Mg}^+ + {\rm C}^+ + {\rm C} + e' \nonumber\\
&\to & {\rm Mg} + {\rm C}^+ + {\rm C}^+ + e''\,.\, \ \ 
\label{MgCC}
\end{eqnarray}
The ionization potential of C amounts to 11.26 eV \cite{NIST}. 
Accordingly, the incident electron needs to have an energy of 
only $\varepsilon_p\ge 0.13$\,eV here.

To estimate the cross section for DICEC in this system, we assume 
an incident electron with energy close to threshold. By using a 
cross section for radiative recombination of such an electron with
Mg$^{2+}$ of $\sigma_{_{\rm RR}}\approx 0.9$\,kb (with the related
photoionization cross section of $\sigma_{_{\rm PI}}\approx 270$\,kb
from \cite{PI-MgII-2}) and a photoionization cross section of C at 
14.87\,eV of $\sigma_{_{\rm PI}}\approx 14$\,Mb \cite{PI-C}, we 
obtain for the first ICEC step in Eq.~\eqref{MgCC} a cross section of 
$\sigma_{_{\rm ICEC}}^{({\rm Mg}^{2+}{\rm C})}\approx 3\times 10^4/R^6$\,a.u.

The second ICEC step is induced by an electron with $\varepsilon_{p'}\ge 3.61$\,eV
(corresponding to a de Broglie wavelength of $\lambda_{p'}\lesssim 2$\,a.u.).
We take a typical cross section value of $\sim 200$\,kb for the (nonresonant) 
photoionization of Mg, which corresponds to a cross section for radiative 
recombination with Mg$^+$ of $\sigma_{_{\rm RR}}\sim 0.003$\,kb. Combined 
with the cross section for photoionization of C close to threshold of 
$\sigma_{_{\rm PI}}\approx 16$\,Mb \cite{PI-C}, we obtain a cross section
for the second ICEC step of $\sigma_{_{\rm ICEC}}^{({\rm Mg}^+{\rm C})}
\sim 4\times 10^2/R^6$\,a.u.

The cross section for the DICEC process in Eq.~\eqref{MgCC} thus becomes
$\sigma_{_{\rm DICEC}}\sim 1.2\times 10^7/(2\pi R^{14})$\,a.u. which 
amounts to approximately 0.1\,kb at $R=7$\,a.u. This value agrees with
the rough estimate for the DICEC cross section given at the end of Sec.~II.

Interestingly, at 11.3--11.5\,eV there are autoionizing $3p4d$ and $3p6s$ 
resonances in Mg, which lead to high cross sections of 3--4 Mb for resonant
photoionization \cite{PI-MgI-2}. Accordingly, for such resonant energies, the 
electron emitted after the first ICEC step can recombine dielectronically 
with the Mg$^+$ ion at a cross section of $\sigma_{_{\rm DR}}\sim 0.1$\,kb. 
In this scenario, the second ICEC step may arise if the 
excited Mg atom stabilzes by transferring the transition energy to the 
neighboring C atom, causing its ionization \cite{res-ICEC1}. The cross section 
for this resonant DICEC can be estimated to amount few kb at $R=7$\,a.u.

The DICEC process of Eq.~\eqref{MgCC} could be realized experimentally
by utilizing endohedral fullerenes, where a Mg$^{2+}$ is located inside
a C$_{60}$ cage. The latter has a radius of about 7\,a.u. \cite{Radius-C60}.
In such a system, the ICEC and DICEC cross sections may benefit from
amplifications by the large configuration number \cite{ICEC-JPB}.
We note in this context that photoionization experiments on endohedral 
fullerenes have successfully been carried out \cite{Exp-C60}. Theoretically, 
also interatomic Coulombic decay (ICD) processes, including double ICD, have 
been studied in various endohedral fullerenes \cite{ICD-C60, ICD-Ar-C60, Radius-C80}.

\medskip
Before moving on to the conclusion, we list some further conceivable 
three-center systems where DICEC may occur.

\medskip

(i) Earth alkaline dication in hydration shell of water:
\begin{eqnarray*}
e + {\rm Mg}^{2+} + {\rm H}_2{\rm O} + {\rm H}_2{\rm O} 
&\to & {\rm Mg}^+ + ({\rm H}_2{\rm O})^+ + {\rm H}_2{\rm O} + e' \\
&\to & {\rm Mg} + ({\rm H}_2{\rm O})^+ + ({\rm H}_2{\rm O})^+ + e''\,.
\end{eqnarray*}
This example extends the corresponding ICEC consideration of \cite{ICEC-JPB}
to double ICEC. Since the ionization potential of water is about 12.62 eV, the 
incident electron needs to have an energy $\varepsilon_p\ge 2.85$\,eV here 
(where the ICEC cross section in the first step is $\lesssim 100$\,kb \cite{ICEC-JPB}).

\medskip

(ii) Earth alkaline dication and halogen anions: 
\begin{eqnarray*}
e + {\rm Ca}^{2+} + {\rm Cl}^- + {\rm Cl}^- 
&\to & {\rm Ca}^+ + {\rm Cl} + {\rm Cl}^- + e' \\
&\to & {\rm Ca} + {\rm Cl} + {\rm Cl} + e''\,.
\end{eqnarray*}
Here, the energy of the incident electron can be arbitrarily small ($\varepsilon_p\ge 0$)
given the electron detachment potential of Cl$^-$ being 3.60 eV.
Since CaCl$_2$ exists as a molecule, the above example might potentially 
arise in solution, wherein CaCl$_2 \to$ Ca$^{2+}$ $+$ 2\,Cl$^-$.

\medskip

(iii) Via double ICEC, a positively charged ion might turn into a negatively charged one, e.g., according to
\begin{eqnarray*}
e + {\rm H}^+ + {\rm Mg} + {\rm Mg} 
&\to & {\rm H} + {\rm Mg}^+ + {\rm Mg} + e' \\
&\to & {\rm H}^- + {\rm Mg}^+ + {\rm Mg}^+ + e''\,.
\end{eqnarray*}
Since the binding energy of H$^-$ is about 0.75\,eV, the minimum 
energy of the incident electron to induce this process is approximately 0.95\,eV.
\medskip

(iv) Finally, an example for Eq.~\eqref{DICEC_3a} would be:
\begin{eqnarray*}
e + {\rm Mg}^+ + {\rm Ca} + {\rm Mg}^+ &\to & {\rm Mg} + {\rm Ca}^+ + {\rm Mg}^+ + e' \\
&\to & {\rm Mg} + {\rm Ca}^{2+} + {\rm Mg} + e''\,.
\end{eqnarray*}
The incident electron needs to have an energy of $\varepsilon_p\ge 2.61$\,eV, 
so that the intermediate electron is ejected with $\epsilon_{p'}\ge 4.15$\,eV.\\

\section{Conclusion}

The two-step process of double ICEC in two- and three-center atomic systems has been studied.
Here, the electron resulting from a first ICEC between centers $A$ and $B$ induces a second 
ICEC with participation of $A$ and (or) $B$. By using an intuitive theoretical approach, the 
cross sections for the various versions of double ICEC have been obtained. While the absolute 
magnitude of the cross section was found to be rather low ($\sigma_{_{\rm DICEC}}\lesssim 1$\,kb),
it must be viewed from the perspective that double ICEC involves in total four bound $\leftrightarrow$ 
free transitions and thus leads to strong changes in the electron configuration of the system. 
Several potential atomic systems where double ICEC can occur, have been discussed. As particularly promising, a three-center system composed of an earth alkaline dication like Mg$^{2+}$ (or Ca$^{2+}$) in the neighbourhood of two carbon atoms has been identified, as could be realized experimentally in the form of endohedral fullerenes.

It should be noted that ICEC, while having been studied by theoreticians in quite some detail,
has not been observed experimentally yet. In light of this, an observation of double ICEC 
appears at present very challenging, given its small cross section. Nevertheless, promising 
prospects towards an experimental test of single ICEC have recently been outlined \cite{ICEC-Review}. 
And several other interatomic processes induced by electron impact have already been observed in experiment \cite{Exp1,Exp2,Exp3,Exp4}. 



\begin{thebibliography}{33}
\bibitem{remark} We recall that the geometrical approach does not apply 
when the energy of the electron, that travels between the centers, is very small.
\end{thebibliography}
\end{document}